\documentstyle[12pt,twoside,fleqn,espcrc1,epsfig]{article}

\newcommand{\AmS}{{\protect\the\textfont2
  A\kern-.1667em\lower.5ex\hbox{M}\kern-.125emS}}
\newcommand{\beq}{\begin{equation}}
\newcommand{\eeq}{\end{equation}}
\newcommand{\beqn}{\begin{eqnarray}}
\newcommand{\eeqn}{\end{eqnarray}}

\newcommand{\lton}{\mathrel{\lower.9ex
                  \hbox{$\stackrel{\displaystyle <}{\sim}$}}}
\def\gsim{\mathrel{\rlap{\lower4pt\hbox{\hskip1pt$\sim$}}
\raise1pt\hbox{$>$}}}
\def\lsim{\mathrel{\rlap{\lower4pt\hbox{\hskip1pt$\sim$}}
\raise1pt\hbox{$<$}}}

\title{Charmonium Production off Nuclei: from SPS to RHIC}

\author{B.Z.~Kopeliovich\address{Max-Planck Institut f\"ur Kernphysik, 
Postfach 103980, 69029 Heidelberg and\\
Institut f\"ur Theoretische Physik der Universit\"at, 93040 Regensburg,
Germany}}

\begin{document}

\maketitle

\begin{abstract}

The physics of charmonium suppression in nuclear collisions drastically
changes between the energies of SPS and RHIC. Mechanisms suppressing
charmonia at the SPS are reviewed, neither of which is important at RHIC.  
On the other hand, coherence, or shadowing of $c$ quarks and gluons
barely seen at SPS, becomes a dominant effect at RHIC providing a much
stronger suppression.  The onset of coherence at Fermilab energies
explains why the observed cross section ratio falls steeply at large
Feynman $x_F$. In nuclear collisions the variation of charmonium
suppression with $x_F$ suggests a sensitive probe to look for the QGP.

\end{abstract}


\medskip\noindent
{\bf 1. Introduction}

\smallskip
 Charmonium production in heavy ion collisions at high energies is
subject to a two-step suppression well separated in time,
 \beq
S^{\Psi}_{AB}=S_{nucl}\ S_{FSI}\ .
\label{10}
 \eeq
 The early stage presented by the first factor is due to multiple
interaction with the colliding nucleons. It includes absorption and
shadowing. At the late stage the formed charmonium attenuates through the
cloud of matter produced in the collision, providing the second
suppression factor in (\ref{10}). Actually, this is the factor probing
the properties of the created matter and is the main goal of measurements
\cite{na50}. However, it is impossible to single the final state
interaction (FSI) factor out of the overall suppression in Eq.~(\ref{10})
unless the first factor, $S_{nucl}$, providing the main contribution, is
reliably known. Surprisingly, this problem is not settled yet even for
$pA$ collisions.

\medskip\noindent
{\bf 2. Pitfalls and perspectives for the QGP search at the SPS}

\smallskip
We give here a brief list of effects contributing to nuclear suppression of 
the charmonium production rate in $pA$ and $AB$ collisions at SPS energies.

{\it 1. Absorption.} The model for the factor $S_{nucl}$ widely used at
the SPS is based on a simple idea that the charmonium is produced
instantaneously inside the nucleus and then attenuates with a constant
absorption cross section treated as a fitted parameter.  This model is
missing important effects and predicts a suppression which is independent
of Feynman $x_F$, while the data show that $S_{nucl}(x_F)$ falls
dramatically with $x_F$ \cite{na3,katsanevas}.

{\it 2. Formation time.} Apparently, the produced $\bar cc$ pair takes
time to develop the charmonium wave function. According to the
uncertainty principle one cannot momentarily disentangle the different
levels and conclude which state has been produced.  It takes a time,
called the formation time (length), which is inversely related to the
charmonium mass splitting. Corrected for Lorentz time dilation it can be
estimated to be $t_f=2E_{\Psi}/(M_{\Psi^\prime}^2-M_{J/\Psi}^2)$. The
NA38/50 experiments \cite{na50} detect charmonia with $E_{\Psi}\approx
50\,GeV$, therefore $t_f\gsim R_A$ and the formation effects are
important. A quantum-mechanical description of the evolution of a $\bar
cc$ wave packet in an absorptive medium was suggested in \cite{kz91}, and
in a hadronic representation in \cite{hk-prl}. A beautiful quantum effect
has been found: in spite of naive expectations the $\Psi'$ attenuation
should be similar or even less than that of $J/\Psi$, provided that the
energy is sufficiently high \cite{kz91}. 
At the same time, in $BA$ collisions, $\Psi'$ is more 
suppressed due to
the inverse kinematics for nuclei $A$ and $B$ \cite{hk-prl}. At the same
time the effective absorption cross for $J/\Psi$ varies with energy and
nuclear thickness and cannot be treated as a universal parameter
\cite{hhk}.

{\it 3. Energy loss.} This mechanism was suggested in \cite{kn} right
after the NA3 collaboration published the first data demonstrating a
steep drop of the cross section ratio at large $x_F$. The incoming hadron
experiences an inelastic collision on the surface of the nucleus,
followed by hadronization and energy loss of the projectile partons. As a
result, they arrive at the point of $\Psi$ creation with diminished
energy. Therefore, the value of $x_F$ in this elementary process must be
shifted to a higher value resulting in extra suppression. A similar, but
weaker nuclear suppression was also predicted in \cite{kn} for the
Drell-Yan reaction, confirmed by the E772/E866 experiments. The recent
analysis \cite{eloss} of this data led to the rate of energy loss per
unit of length $dE/dz=-2.3\pm0.52\pm 0.5\,GeV/fm$ in agreement with the
value used in \cite{kn}. The mechanism of energy loss alone is able to
describe the main features of the NA3 data, later confirmed by the E537
experiment \cite{katsanevas}.

{\it 4. Gluon enhancement in nuclei.} There is some model-dependent
evidence \cite{antishad} that the gluon density is enhanced in heavy
nuclei by about $10-20\%$ at large $x_2\sim 0.1$. Accidentally, this is
just the value corresponding to the kinematics of the NA38/50
experiments. Therefore, theoretical predictions should be corrected for
this factor $\sigma_{pA}^{\Psi}/\sigma_{pN}^{\Psi} \propto
G_A(x_2)/G_N(x_2) \approx 1.1-1.2$.

{\it 5. Excitation of nuclear matter in $AB$ collisions.} It is commonly
assumed that charmonium produced in heavy ion collisions propagates and
attenuates in conventional cold nuclear matter. However, each nucleon met
by the $\Psi$ has already interacted at least once with other nucleons
and must be in an excited state. First of all, inelastic $NN$ collisions
are followed by gluon radiation which also contributes to the break-up of
the $\Psi$ \cite{gluons}. Second, inelastic $NN$ collisions are mediated
by color exchange. Therefore, all the nucleons met by the $\Psi$ are in
colored states and interact stronger than colorless ones \cite{hkp}.
These effects are able to explain the observed suppression of the total
cross section of charmonium production, including lead-lead, but do not
leave much room for the FSI suppression which manifests itself in central
collisions at large $E_T$.

\begin{figure}[htb] \begin{minipage}[t]{70mm} \vspace*{-150pt}
\rotatebox{270}{\includegraphics[width=45mm]{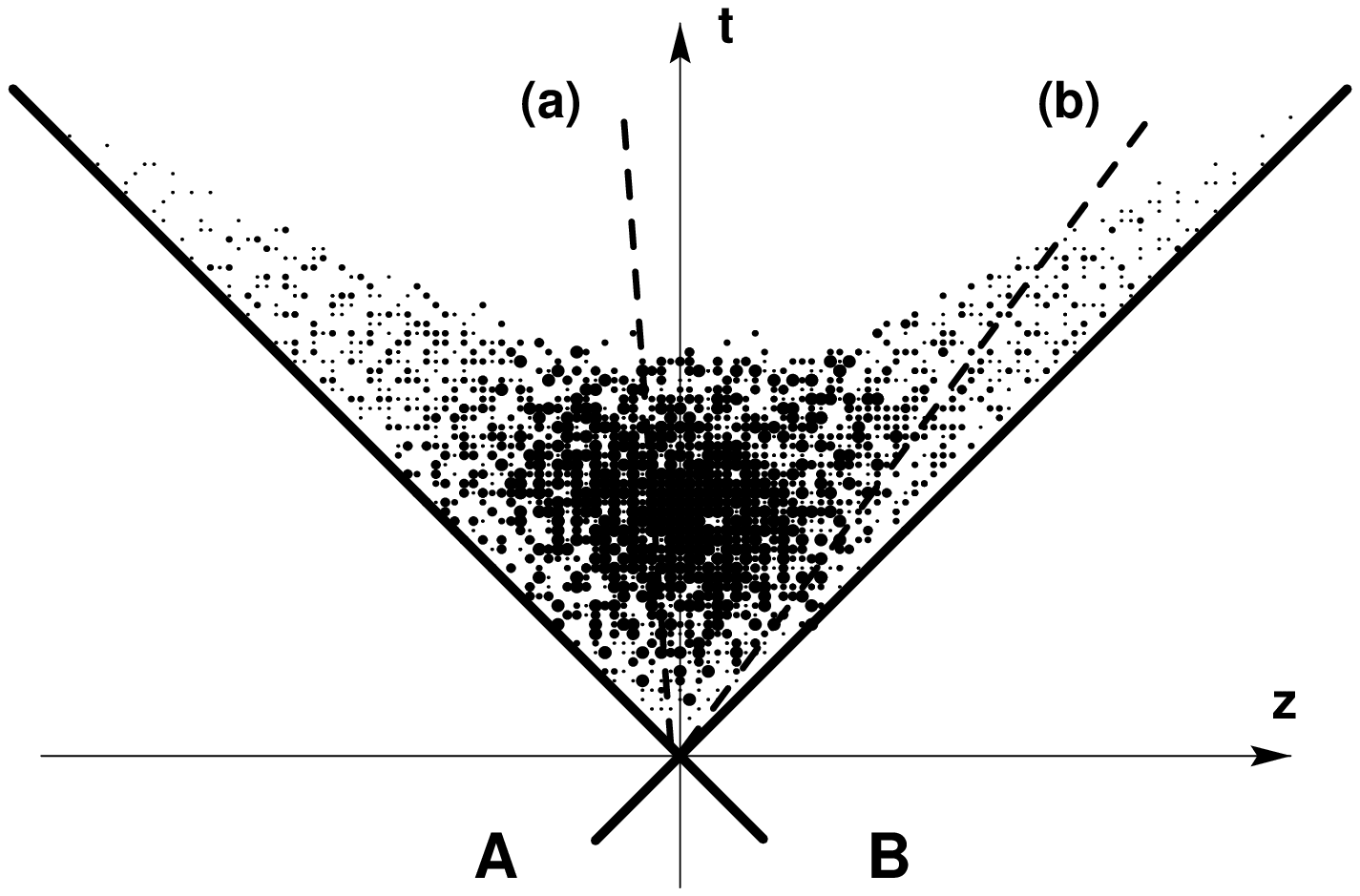}} \vspace{5mm}
\caption{Development of a nucleus-nucleus collision as function of
longitudinal coordinate $z$ and time $t$ in the c.m.  frame. Dashed lines
show trajectories of charmonia produced with small (a) and large (b)
values of $x_F$.} \label{medium}
 \end{minipage}
 \hspace{\fill}
 \begin{minipage}[t]{80mm}
\includegraphics[width=80mm]{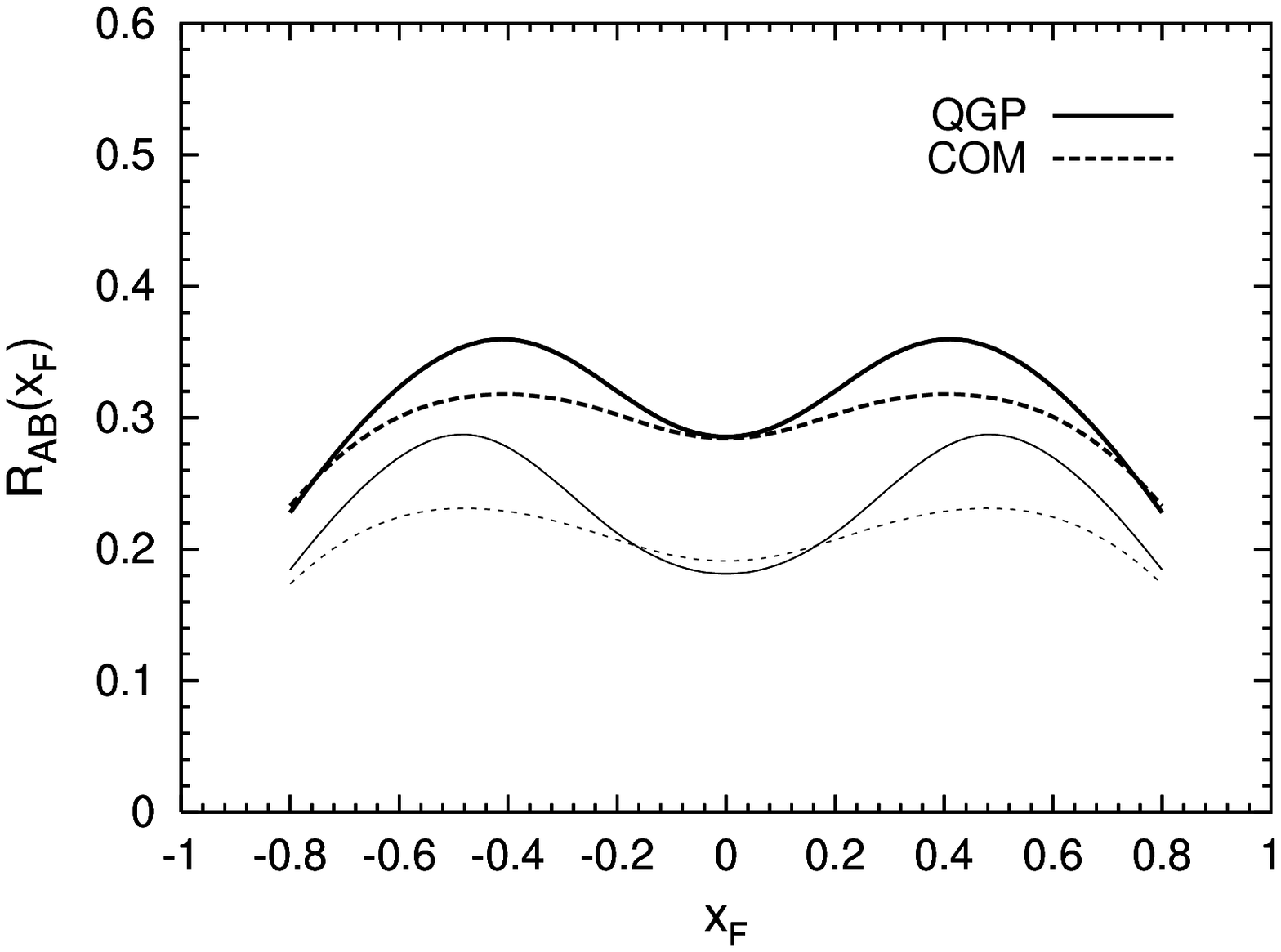}
\vspace{-10mm}
 \caption{Cross sections of charmonium production as function of $x_F$ at
SPS.  Full curves refer to the plasma model, while dashed to comovers.
Thick and thin curves correspond to the total cross section and central
collisions respectively.}
 \label{sps}
\end{minipage}
 \end{figure}

{\it 6. Scanning the QGP.} Although the observed $E_T$ dependence of
charmonium suppression cannot be reproduced by simplest models, still
more sophisticated approaches are more successful, and manifestation of
the FSI suppression is disputable (e.g.  see \cite{q}). A new sensitive
probe for FSI, scanning the produced matter by varying $x_F$ of the
charmonium, has been suggested recently \cite{hkp2}. One can see it on
the plot Fig.~\ref{medium} illustrating the time -- longitudinal
coordinate distribution of the created medium.
 Interaction with the medium most dense at central rapidities causes a
specific minimum at $x_F=0$ in the nuclear suppression factor. This would
be an indisputable signature of FSI, which is also very sensitive to QGP
production.

\medskip\noindent
{\bf 3. From Fermilab to RHIC.} 

\begin{figure}[htb]
\begin{minipage}[t]{75mm}
\vspace*{-8cm}
\includegraphics[width=80mm]{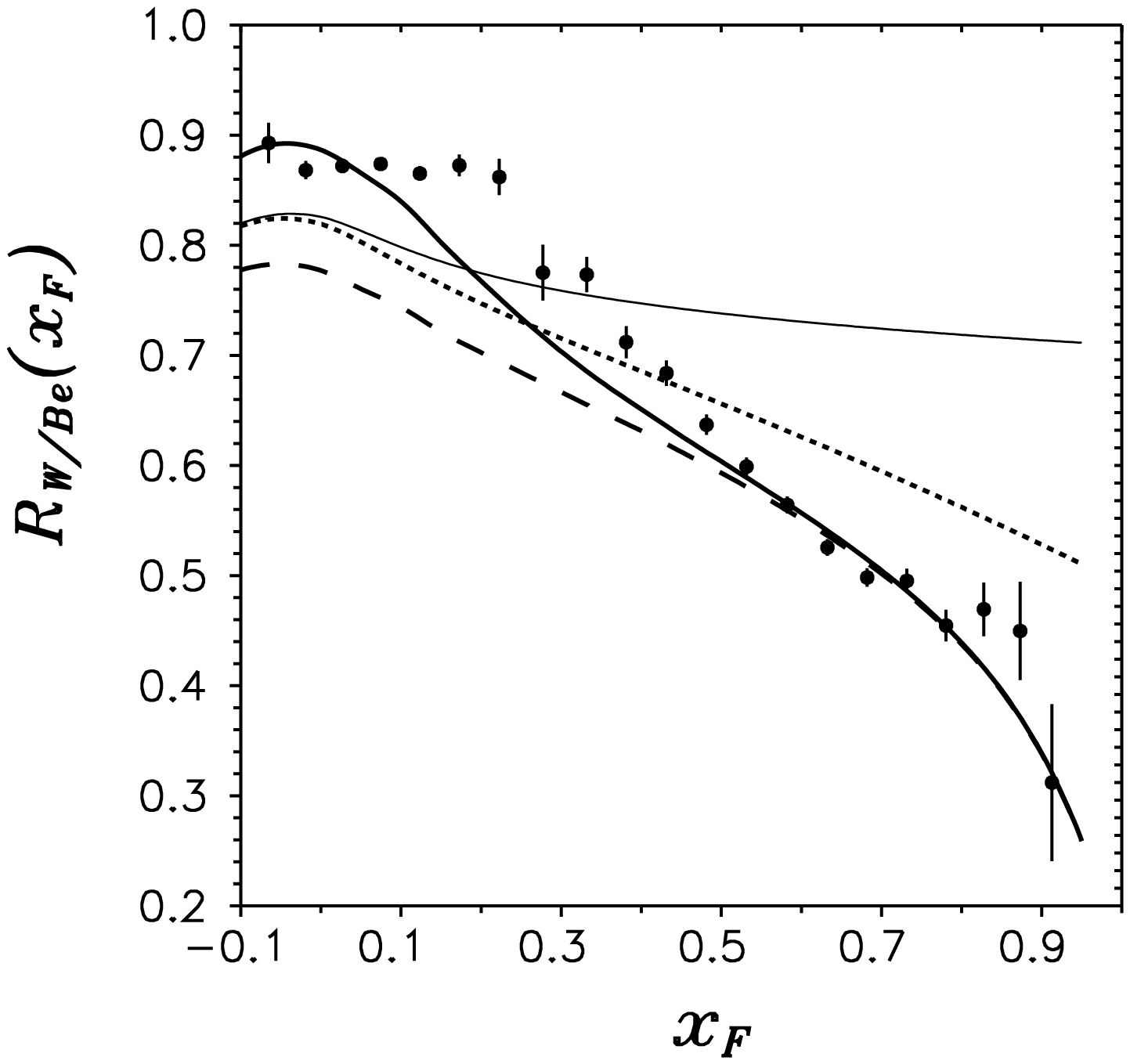}
\vspace{-3cm}
 \caption{Comparison of the results of parameter free calculations in
\cite{kth} with the E866 data \cite{e866}. See text for explanations for
the curves.}
 \label{fnal}
 \end{minipage}
 \hspace{\fill}
 \begin{minipage}[t]{75mm}
\includegraphics[width=80mm]{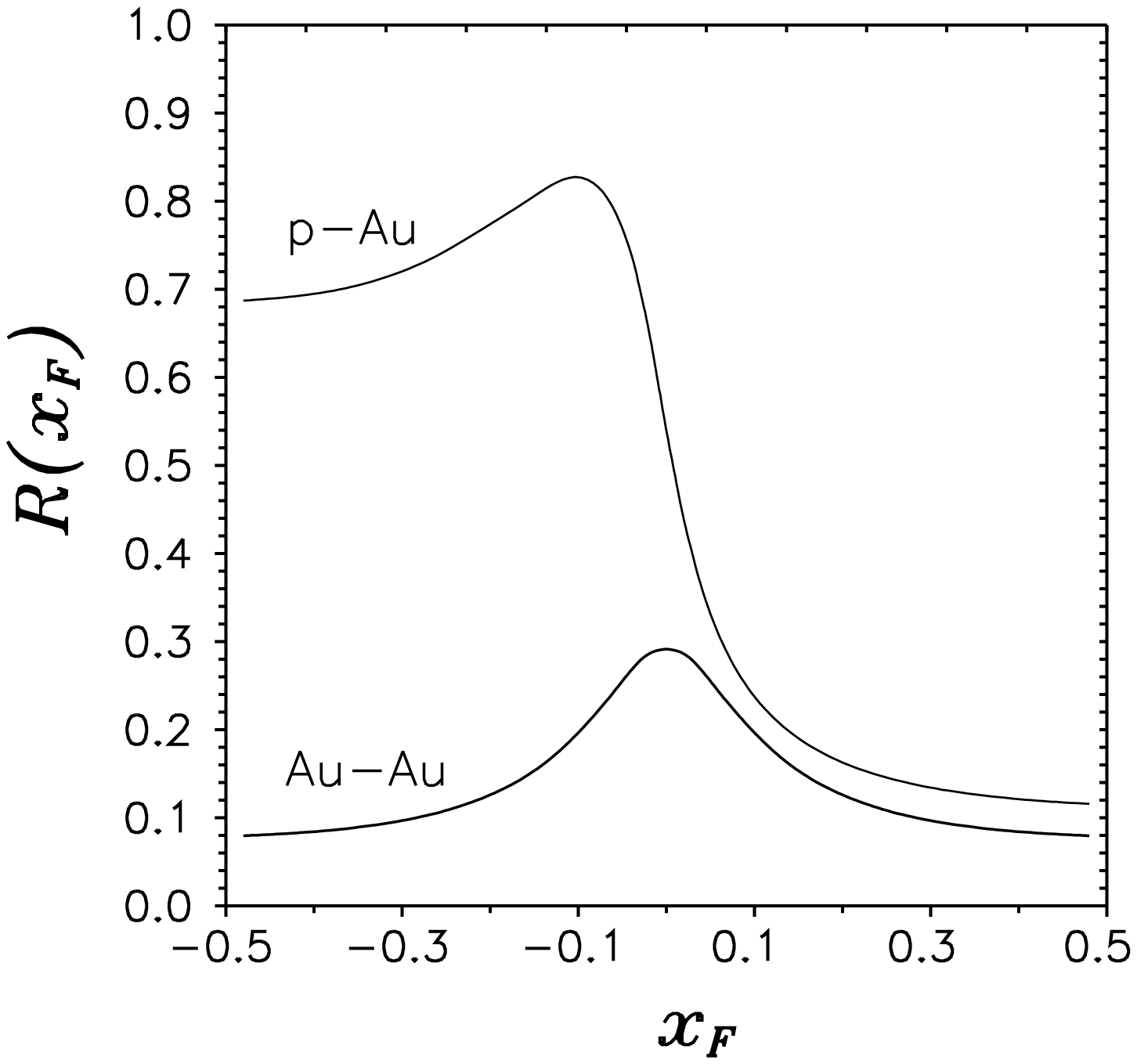}
\vspace{-3cm}
 \caption{Predictions of \cite{kth} for the $x_F$ dependence of nuclear
suppression in proton-gold and gold-gold collisions at
$\sqrt{s}=200\,GeV$.}
 \label{rhic}
\end{minipage}
 \end{figure}

\smallskip
 A new phenomenon, quantum coherence, or quark and gluon shadowing is
expected to become dominant at RHIC. However, before jumping to RHIC
energies, one must understand (as observed at Fermilab \cite{e866}) the
$x_F$ dependence of $Psi$ suppression, which demonstrates the onset of
coherence. The first full QCD (parameter free) calculation of the
dependence of nuclear suppression on $x_F$ is performed in \cite{kth}.
The results are compared with data \cite{e866} in Fig.~\ref{fnal} where
contributions of different mechanisms are explicitly shown.
 The curves correspond to:  absorption and shadowing for the produced
$\bar cc$ (thin solid curve); gluon shadowing added (dotted); energy loss
included (dashed); gluon antishadowing added (thick solid). One can see
that gluon shadowing is the main source of nuclear suppression at small
$x_2$.

Energy loss violating $x_2$ scaling still noticeable at Fermilab
completely vanishes at RHIC. Predictions \cite{kth} for nuclear
suppression in $p-Au$ and $Au-Au$ collisions at RHIC are depicted in
Fig.~\ref{rhic}. On top of that one should expect a FSI suppression which
is strongest at $x_F=0$.

An opposite effect of FSI enhancement of $\Psi$ due to fusion of produced
$\bar cc$'s has been predicted recently \cite{bs,thews}. However, if
gluon shadowing suppresses direct $\Psi$ by nearly an order of magnitude,
it should be square of that for the fusion mechanism. It may eliminate
fusion even at LHC due to gluon saturation. Apparently, this problem
needs further study. Nevertheless, whatever happens, FSI suppression or
enhancement, it will show up only on top of the nuclear suppression which
has to be well understood.

{\bf Acknowledgement:} I am thankful to J.~H\"ufner, A.~Polleri and
A.V.~Tarasov for fruitful and enjoyable collaboration. I am also grateful
to Tim Hallman for careful reading the manuscript and making numerous
improving comments. Partial support from the GSI grant No.~GSI-OR-SCH is
acknowledged.

\end{document}